\documentclass[unnumsec,webpdf,contemporary,large]{oup-authoring-template}

\graphicspath{{Fig/}{figures/}}

\usepackage{booktabs}
\usepackage{multirow}
\usepackage{array}
\usepackage{amsmath}
\usepackage{amssymb}
\usepackage{siunitx}
\usepackage{xspace}
\usepackage{url}
\usepackage{listings}
\usepackage{xcolor}
\usepackage{caption} 
\usepackage{placeins}

\newcommand{\toolname}{\textsc{AcmGENTIC}\xspace}
\newcommand{\acmg}{ACMG/AMP\xspace}
\newcommand{\psthree}{PS3\xspace}
\newcommand{\bsthree}{BS3\xspace}

\begin{document}

\journaltitle{}
\DOI{}
\copyrightyear{}
\pubyear{}
\vol{}
\issue{}
\access{}
\appnotes{}

\firstpage{1}

\title[LLM for Functional Evidence Extraction]{Large Language Models for Variant-Centric Functional Evidence Mining}

\author[1,2]{Ali Saadat}
\author[1,2,3,$\ast$]{Jacques Fellay}

\address[1]{\orgdiv{School of Life Sciences}, \orgname{Ecole Polytechnique Fédérale de Lausanne},\orgaddress{ \state{Lausanne}, \country{Switzerland}}}

\address[2]{\orgdiv{Swiss Institute of Bioinformatics}, \orgaddress{\state{Lausanne}, \country{Switzerland}}}

\address[3]{\orgdiv{Precision Medicine Unit, Biomedical Data Science Center}, \orgname{Lausanne University Hospital and University of Lausanne}, \orgaddress{\state{Lausanne}, \country{Switzerland}}}

\corresp[$\ast$]{Corresponding author: \href{mailto:jacques.fellay@epfl.ch}{jacques.fellay@epfl.ch}}

\abstract{%
Functional evidence is central to clinical interpretation of genomic variants, yet linking variants to the relevant papers and translating experimental results into usable evidence statements remain labor-intensive. We constructed a benchmark anchored to ClinGen-curated annotations and evaluated two large language models (LLMs), including a non-reasoning model (gpt-4o-mini) and a reasoning model (o4-mini), on the following tasks: 1)  abstract screening: given a paper abstract, whether the study reports a functional experiment that directly tests specific variants. 2) full-paper evidence extraction and classification: given the full PDF of a matched variant--paper pair, to extract the key experimental readouts and interpret the direction of evidence for clinical curation and an associated evidence summary. Starting from the full set of ClinGen-curated variants, we selected those annotated with functional evidence. We then processed curator comments with an LLM to extract PubMed identifiers, evidence labels, and relevant narrative, and used these identifiers to retrieve titles, abstracts, and open-access PDFs, yielding variant--paper pairs for evaluation. For abstract screening, both models achieved high recall (0.88--0.90) with moderate specificity (0.59--0.65). For full-text evidence classification under an explicit variant-matching gate, o4-mini achieved 96\% accuracy and substantially higher specificity (0.83 vs.\ 0.37) while maintaining high F1 (0.98 vs.\ 0.96) compared with gpt-4o-mini. We additionally employed an LLM-as-judge protocol to compare LLM-generated evidence summaries against expert-written curator comments. Finally, we implemented \toolname, an end-to-end pipeline that takes variant coordinates, expands them via alternate identifiers and synonyms, retrieves candidate literature via LitVar2, filters abstracts with LLMs, acquires PDFs, performs multimodal extraction using an LLM-powered workflow, and generates an evidence report for curator review. \toolname additionally supports an optional agentic extraction mode for deeper parsing of figures and tables, which we include as an extensibility feature. Overall, this benchmark and pipeline provide a practical foundation for scaling functional-evidence curation through human-in-the-loop LLM assistance.%
}


\keywords[Abbreviations]{LLM, AI, GPT, ACMG, AMP, ClinGen, VEP, OCR, PDF}

\boxedtext{Key Messages}{
\begin{itemize}
\item A ClinGen-derived benchmark for functional evidence extraction at abstract- and full-paper levels.
\item Under a variant-matching gate, the reasoning model (\texttt{o4-mini}) achieves 96.3\% accuracy and markedly higher specificity (0.828 vs.\ 0.371) than the non-reasoning model (\texttt{gpt-4o-mini}) for \psthree/\bsthree direction classification.
\item \toolname is an open-source variant-to-report pipeline that takes a genetic variant coordinates and collects all the literature reported functional experiments and their conclusion in a user-friendly format which can be reviewed further by human experts.
\end{itemize}
}

\maketitle

\section{Introduction}

Clinical interpretation of genomic variants requires integrating heterogeneous evidence into a standardized classification of pathogenicity. The \acmg guidelines formalize this process through criteria spanning population data, segregation, de novo occurrence, computational predictions, and functional studies~\citep{Richards2015-xk}. Functional evidence plays a distinctive role because it can provide mechanistic support directly linking a molecular perturbation to a disease-relevant effect. Within this framework, \psthree denotes well-established functional studies supportive of a damaging effect, whereas \bsthree denotes well-established functional studies supportive of a lack of damaging effect~\citep{Brnich2019-zm}.

Despite its value, functional evidence remains difficult to use at scale. Variant mentions are inconsistent across the literature (rsIDs, HGVS strings across transcripts \citep{Hart2024-nn}, protein-level shorthand, or legacy nomenclature), complicating variant-level retrieval and identity alignment. Moreover, experimental details needed for clinical interpretation are often absent from abstracts and distributed across the text, figures, and tables. ClinGen~\citep{Rehm2015-he} recommendations clarify how assay validity, calibration, and concordance with disease mechanism should affect \psthree/\bsthree application and strength (supporting, moderate, strong, very strong), yet these signals are rarely expressed in a uniform, machine-readable form~\citep{Brnich2019-zm}.

Recent work has begun to operationalize large language models (LLMs) for evidence-centric clinical genetics workflows. CGBench introduced a ClinGen-derived benchmark that tests whether language models can extract and score evidence from scientific papers under guideline-style instructions, including judge-based evaluation against curator explanations~\citep{Queen2025-pr}. The Evidence Aggregator (EvAgg) demonstrated a generative-AI pipeline for rare disease case analysis that retrieves gene-relevant publications and extracts structured case and variant details for analyst review~\citep{Twede2025-db}. AutoPM3 targeted a specific ACMG/AMP evidence category (PM3), combining retrieval with structured extraction to identify supporting evidence from both narrative text and tables~\citep{Li2025-vx}. Together, these efforts highlight both the promise and the challenges: variant interpretation requires scalable literature screening, robust multimodal understanding, and human-centered outputs that present evidence as traceable statements, so curators can rapidly verify specific claims.

\begin{figure}[t]
\centering
\includegraphics[width=\linewidth]{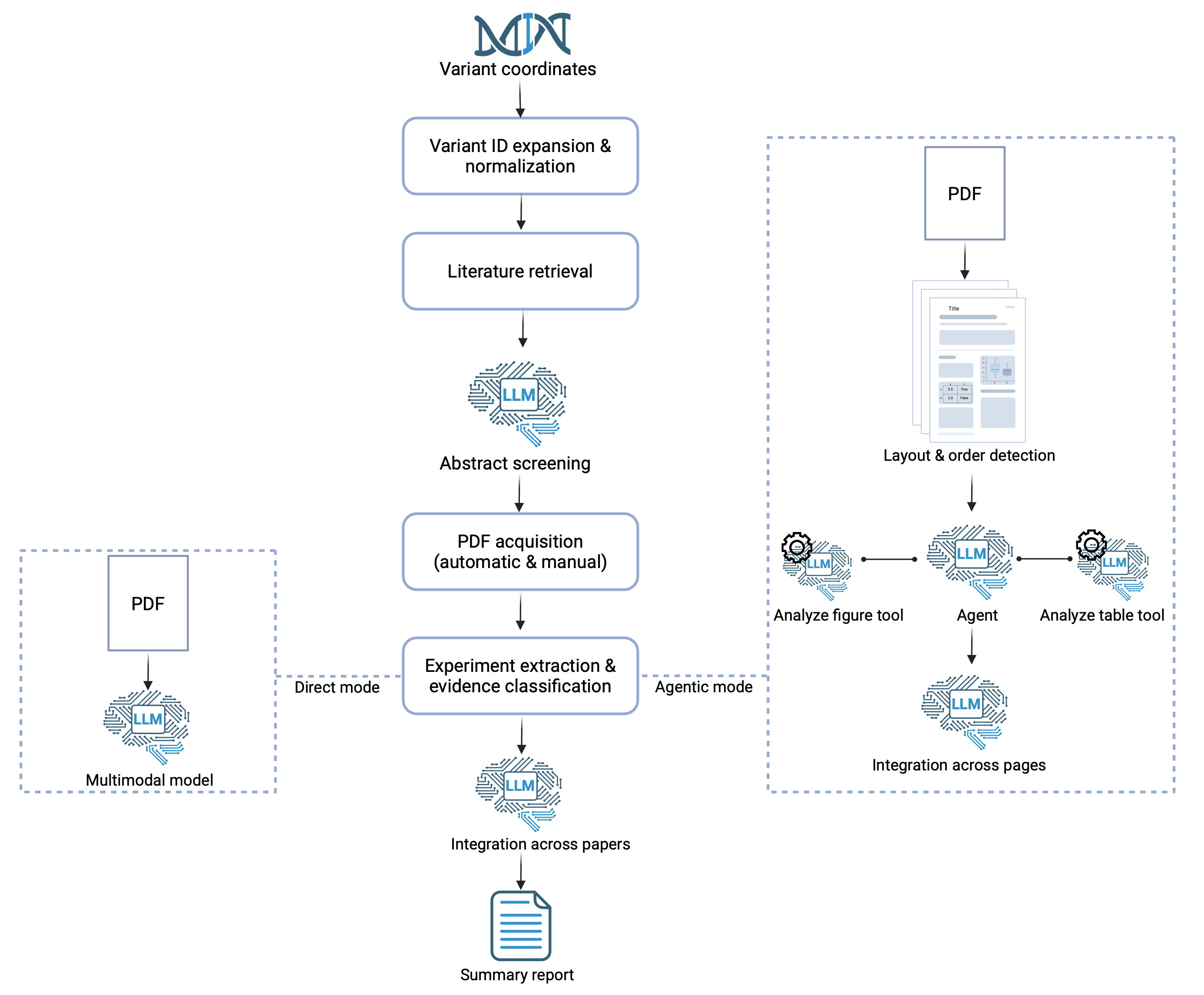}
\caption{\textbf{Overview of \toolname.} Starting from a user-specified variant, \toolname performs (i) variant normalization and synonym expansion across common identifiers, (ii) literature retrieval and abstract-level screening to prioritize likely functional studies, (iii) PDF acquisition and variant matching, (iv) structured experiment extraction and \psthree/\bsthree-aligned evidence interpretation, and (v) report generation for curator verification and downstream use.}
\label{fig:acmgentic-overview}
\end{figure}

Here, we focus on variant-centric functional evidence mining for \psthree/\bsthree. We (i) construct a ClinGen-anchored benchmark for abstract screening and full-paper functional evidence extraction; (ii) evaluate two multimodal LLMs with different reasoning behavior under an explicit variant-matching gate to reduce misattribution risk; and (iii) implement \toolname, an end-to-end variant-to-report pipeline designed for human-in-the-loop curator review. The empirical results in this paper leverage \toolname's direct extraction mode, though the system also supports an agentic mode which demonstrates its extensibility. Figure~\ref{fig:acmgentic-overview} summarizes the overall \toolname workflow.

\section{Materials and Methods}

\subsection{ClinGen curated variants}
We downloaded 11{,}527 ClinGen curated variants \citep{clingen_ev_rep} and filtered to 1{,}709 variants annotated with \psthree\ or \bsthree\ at any strength (supporting, moderate, strong, or very strong). For each selected variant, we used the expert-written comments which explains the available evidences and rationale for classifying a variant into pathogenic, benign, or unknown significance.

\subsection{Structuring summaries and retrieving linked literature}
We transformed curator summaries into structured fields using \texttt{gpt-4.1-mini} via the OpenAI API \citep{openai_api_docs}. The prompt instructed the model to extract PubMed IDs, PS3/BS3 level assignments, strength of the evidence (very strong, strong, moderate, or supporting), and PS3/BS3-relevant narrative which usually explains what conclusions where drawn from certain publications. The full prompt is provided in Appendix~\ref{app:prompts}.

Using the extracted PubMed IDs, we retrieved titles and abstracts programmatically with the \texttt{metapub} \citep{metapub_ref} Python library. When open-access is available, PDFs were downloaded and cached locally. For benchmarking, we restricted evaluation to programmatically retrievable PDFs to support reproducibility. For \toolname end-to-end pipeline, we additionally support manual addition of PDFs to the local cache for papers that cannot be retrieved automatically (e.g., publisher-restricted content).

\subsection{Variant normalization and synonym expansion}
Variants are referenced in heterogeneous formats across databases and articles (HGVS strings, rsIDs, protein changes, or coordinates). To support variant-level retrieval and matching, we constructed synonym sets.

To overcome such inconsistencies, we used VariantValidator \citep{Freeman2024-dl} and Ensembl VEP API \citep{Yates2015-kh} to generate the following identifiers for each variant:
\begin{itemize}
\item rsIDs (when available),
\item HGVS genomics, coding, and protein expressions (HGVSg, HGVSc, and HGVSp) in both 1-letter and 3-letter amino acid notation,
\item genomic coordinates in both GRCh38 and GRCh37,
\item gene symbols.
\end{itemize}

\subsection{Benchmark tasks and datasets}
We designed and evaluated two relevant tasks: abstract-level screening and full-paper evidence extraction/classification.

\subsubsection{Abstract-level functional experiment screening}
The input was a paper abstract. The LLM was instructed to return a binary label indicating whether the abstract describes a wet-lab functional experiment that directly tests one or more genetic variants. We performed this task with two different prompts, one with more standard instructions and one biased toward positive classification to increase sensitivity in order to not miss any papers with relevant experiments.

The dataset consisted of:
\begin{itemize}
\item \textbf{Positive samples:} 529 abstracts from papers linked to ClinGen variants annotated with \psthree/\bsthree.
\item \textbf{Negative samples:} 529 abstracts randomly sampled from papers cited in ClinGen curator comments for variants with no \psthree/\bsthree annotation.
\end{itemize}

\subsubsection{Full-paper experiment extraction and classification}
The input was the full-paper PDF together with a set of identifiers for the target variant, including the gene symbol, rsID, and HGVS descriptions at the genomic, cDNA, and protein levels (with both one-letter and three-letter amino-acid notation). The LLM was instructed to return a structured record comprising:
\begin{itemize}
\item \textbf{Variant matching}: The process of identifying whether variants mentioned in the paper correspond to the target variant by building an equivalents set (same rsID, genomic coordinates, cDNA change, or protein change). The \texttt{match status} indicates the strength of this matching: \emph{matched} (exact match via rsID, genomic, cDNA, or protein notation), \emph{heuristic matching} (plausible match using non-standard notation, e.g., same amino-acid substitution written differently like "R158W mutant" or "R158→W"), \emph{single-variant-study matching} (the paper tests only one specific variant in the gene, with no other variants in the functional results), or \emph{unsuccessful} (no plausible variant found). The \texttt{match type} indicates the identifier used: rsID, genomic, cDNA, protein, multiple (multiple identifiers match), or heuristic. For each match, the system reports \texttt{confidence} (high/medium/low), the \texttt{matched strings}, and brief \texttt{notes} explaining the decision.
\item \textbf{Experiments}: for each relevant experiment, the details are extracted including the assay type, experimental system, material source, readout (with units when available), the normal comparator, result direction (functionally abnormal, functionally normal, intermediate, mixed, or unclear), effect size or statistics when reported, controls and validation, the authors' conclusion, the supporting text location, caveats, and confidence that the experiment pertains to the target variant.
\item \textbf{Overall evidence}: an aggregate PS3/BS3 direction (or not clear), a strength assignment (very strong, strong, moderate, supporting, or not clear), and a brief rationale.
\item \textbf{Summary}: a short narrative (2--5 sentences) explaining the evidence underlying the final decision.
\end{itemize}

We used \emph{not clear} as an explicit no-decision outcome when either the variant identity could not be confidently aligned to the target or the direction and/or strength of functional evidence was not determinable from the text. We report \emph{coverage} as the fraction of examples for which the system made a decision.

\subsection{End-to-end pipeline implementation}
We implemented \toolname, an end-to-end pipeline that takes a variant (chromosome, position, reference allele, alternate allele) and produces a structured evidence report. The pipeline includes:
\begin{enumerate}
\item \textbf{annotation}: Query Ensembl VEP REST API (GRCh38 or GRCh37) to obtain rsID, HGVSc, HGVSp, gene symbol, MANE Select transcript \citep{Morales2022-fb}, and Ensembl transcript ID.
\item \textbf{Literature retrieval}: Query the LitVar2 API \citep{Allot2023-ug} using the rsID to retrieve PubMed IDs of papers mentioning the variant.
\item \textbf{Paper details}: Fetch titles and abstracts from PubMed using Metapub.
\item \textbf{Abstract screening}: Filter papers using an LLM to retain papers likely to contain variant-level functional experiments. The pipeline supports multiple providers: OpenAI (e.g., \texttt{gpt-4o-mini}, \texttt{gpt-4o}), Anthropic (e.g., \texttt{claude-3-5-sonnet}), and Google (e.g., \texttt{gemini-1.5-flash}, \texttt{gemini-1.5-pro}).
\item \textbf{PDF acquisition}: For the abstracts predicted to report functional evidence, attempt automatic PDF download via Metapub; optionally allows the user for manual paper addition to the cache when programmatic retrieval fails.
\item \textbf{Full-text extraction}: Extract evidence from PDFs using either:
\begin{itemize}
\item \textbf{Direct mode (default):} Submit the full PDF to a multimodal LLM in a single call and request variant-specific functional evidence extraction and classification. This mode is relatively fast, but it might struggle to capture certain details that are embedded in visual elements such as figures and plots.

\item \textbf{Agentic mode:} Decompose the paper into page-level units and run a lightweight document-understanding pipeline before calling the LLM. Concretely, the system (i) renders pages to images, (ii) applies optical character recognition (OCR) using PaddleOCR \citep{cui2025paddleocr30technicalreport} or EasyOCR \citep{easy_ocr_ref} to recover machine-readable text and bounding boxes, (iii) infers reading order and section structure (e.g., multi-column layouts, figure/table captions, footnotes) using a layout-aware model (LayoutReader; LayoutLMv3-based \citep{Pang_Faster_LayoutReader_based_2024}), (iv) routes non-textual regions to specialized vision tools for structured parsing (e.g., \texttt{AnalyzeTable} for table grids and cell content, \texttt{AnalyzeChart} for plotted values and axes), and (v) an LLM is used to aggregate the extracted information per page. These components are orchestrated by a LangChain \citep{LangChain_ref} agent that selects tools per page/region, iterates when extraction is incomplete (e.g., low-OCR-confidence blocks or ambiguous captions), and produces a normalized, citation-backed intermediate representation (sections, paragraphs, figures, tables, and linked captions) suitable for downstream evidence interpretation. This mode is slower, but enables more thorough analysis of the paper particularly visual elements such as figures, tables, and charts.

All benchmark results in this paper use the Direct mode; the optional agentic mode is a prototype included to illustrate extensibility and to target figure/table-heavy publications.

\end{itemize}

\item \textbf{Evidence integration}: Aggregate extracted experiments across papers to determine overall \psthree/\bsthree direction, strength, and confidence, guided by \acmg criteria \citep{Richards2015-xk} and ClinGen recommendations \citep{Brnich2019-zm}.

\item \textbf{Report generation}: Generate an HTML report (optionally exported to PDF) summarizing variant metadata, retrieved papers, extracted experiments, and the integrated assessment. For programmatic use, the system can also return a structured JSON (e.g., easily readable as Python dictionaries) containing per-paper extractions and the final decision.
\end{enumerate}

\subsection{Models and evaluation}
We used \texttt{gpt-4.1-mini} to preprocess ClinGen \emph{Summary of interpretation} comments (PubMed ID extraction and functional-evidence cue extraction). For benchmarking, we evaluated:
\begin{itemize}
\item \texttt{gpt-4o-mini}: an efficient multimodal model (non-reasoning);
\item \texttt{o4-mini}: a multimodal reasoning model.
\end{itemize}

All models were accessed via the OpenAI API. We enforced machine-readable responses using Pydantic schemas \citep{pydantic_ref} and orchestrated the LLMs and agents with the LangChain framework.

To assess whether an LLM-generated evidence summary faithfully matched the corresponding ClinGen rationale, we used \texttt{gpt-4.1} as an independent rater \citep{Zheng2023-gl}. The judge was shown (i) the ClinGen expert-written text (only the part relevant to \psthree/\bsthree) and (ii) the LLM-extracted evidence summary, and asked to score their correspondence on a 1--5 ordinal scale (1 = poor match, 5 = near-complete match), emphasizing factual consistency and coverage of key experimental outcomes while penalizing unsupported claims. To reduce stochasticity and mitigate presentation bias, we repeated scoring three times per example with A/B order swapping and aggregated scores by majority vote \citep{Wang2023-zk,Saha2025-kb}.

We report accuracy, precision, recall, F1, and specificity for binary tasks; macro-averaged metrics for multi-class tasks; and \emph{coverage} as the fraction of examples where the model produced a decision (i.e., not \texttt{not\_clear}).

\section{Results}

\subsection{Benchmark assembly and evaluation subsets}
From 11{,}527 ClinGen curated variants, 1{,}709 carried \psthree/\bsthree evidence at any strength. Using curator-linked PubMed IDs and restricting to programmatically retrievable PDFs, we assembled 529 ClinGen-positive variant--paper pairs. After VariantValidator normalization (MANE Select) and identifier consistency checks, 466 pairs remained for full-paper evaluation (63 were excluded due to normalization failures). For abstract screening, we evaluated 1{,}058 abstracts (529 positive, 529 negative).

\subsection{Abstract-level functional experiment screening}
We prompted the LLMs to screen titles and abstracts for whether the paper reports variant-linked functional experiments. As shown in Figure~\ref{fig:abstract} (detailed in Appendix Table~\ref{tab:abstract-highrecall}), both models achieved high recall (sensitivity), consistent with using abstracts as a first-pass filter where missed relevant studies are costly. Recall reached 0.904 for \texttt{gpt-4o-mini} and 0.885 for \texttt{o4-mini}, indicating that both models reliably surfaced most functional studies from abstract-only information, with remaining errors driven by the limited specificity of abstracts for confirming experimental relevance.

\begin{figure}[t]
\centering
\includegraphics[width=0.92\linewidth]{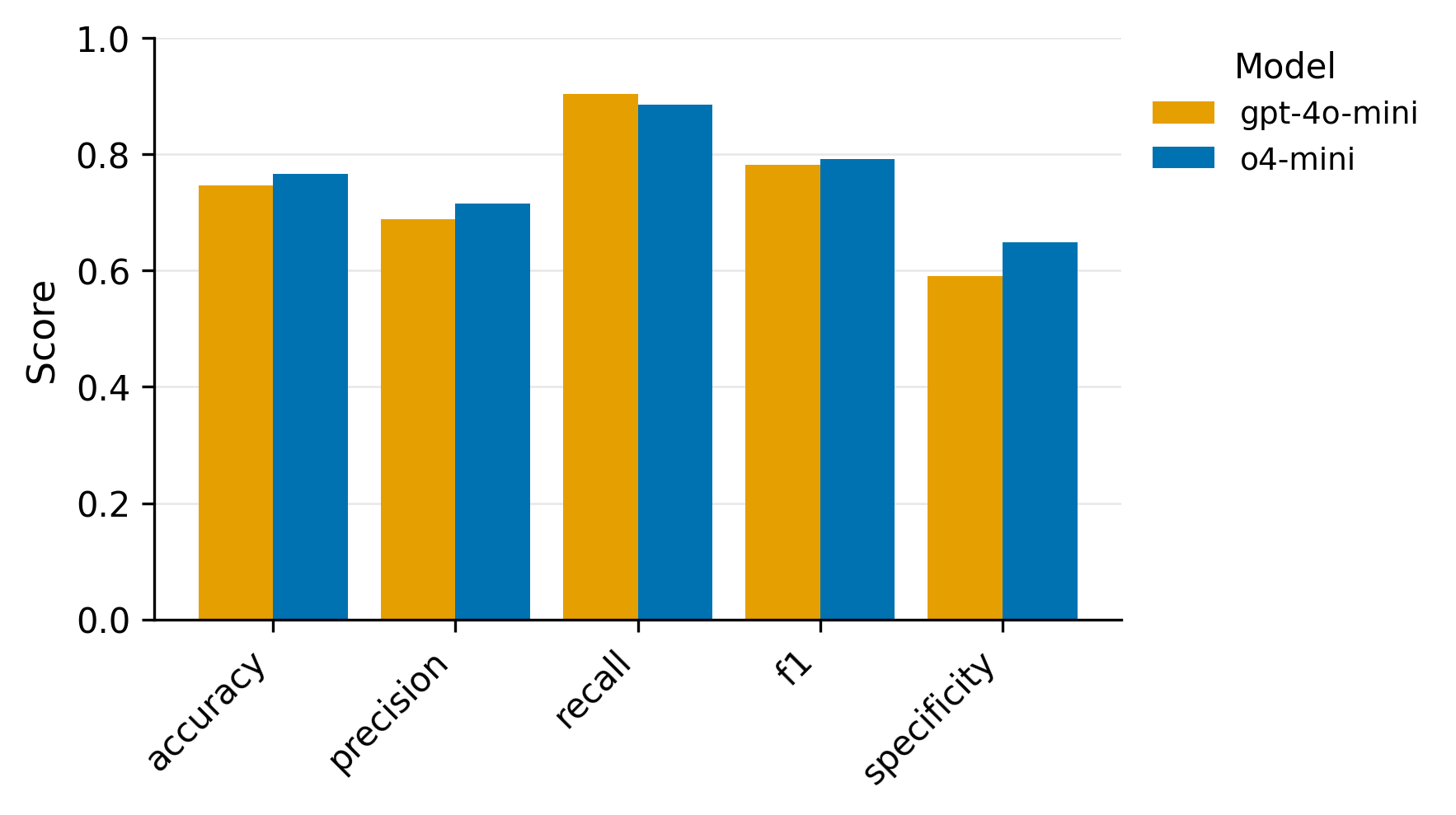}
\caption{Abstract-level screening performance for identifying variant-linked functional experiments (Figure~\ref{fig:abstract}; Table~\ref{tab:abstract-highrecall}). Bars report accuracy, precision, recall (sensitivity), F1, and specificity for \texttt{gpt-4o-mini} and \texttt{o4-mini}. Both models achieve high recall, supporting use as an abstract triage filter to surface candidate functional studies for downstream full-text review.}
\label{fig:abstract}
\end{figure}

\subsection{Variant matching as a bottleneck for full-paper processing}
\begin{figure}[t]
\centering
\includegraphics[width=0.92\linewidth]{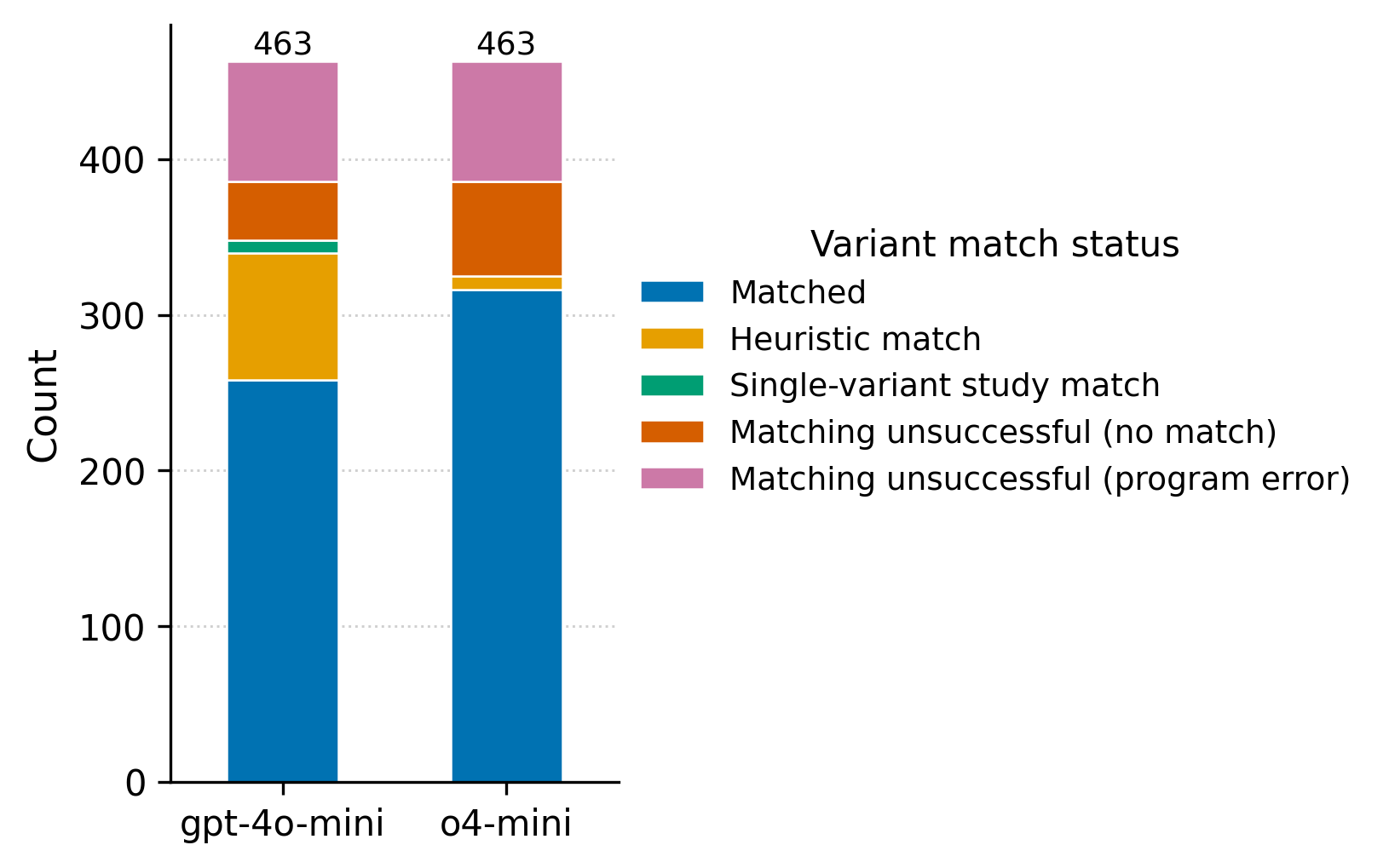}
\caption{Variant matching outcomes by model. Most pairs are successfully matched via exact identifier detection (\texttt{matched}), with additional matches obtained through heuristic alignment or single-variant-study inference. A substantial fraction remains unmatched.}
\label{fig:variant-matching}
\end{figure}

Full-paper extraction is only meaningful when the paper can be linked confidently to the target variant. Figure~\ref{fig:variant-matching} (Appendix Table~\ref{tab:variant-matching}) shows that successful matching was achieved for 339 examples with \texttt{gpt-4o-mini} and 323 with \texttt{o4-mini}. A substantial minority (26--30\%) were labeled \texttt{variant\_matching\_unsuccessful}. Given the clinical risk of attributing functional findings to the wrong variant, the system is designed to prefer \texttt{not\_clear} (no-decision) when alignment is uncertain.

\subsection{Full-text \psthree\ vs.\ \bsthree direction classification}
\begin{figure}[t]
\centering
\includegraphics[width=0.92\linewidth]{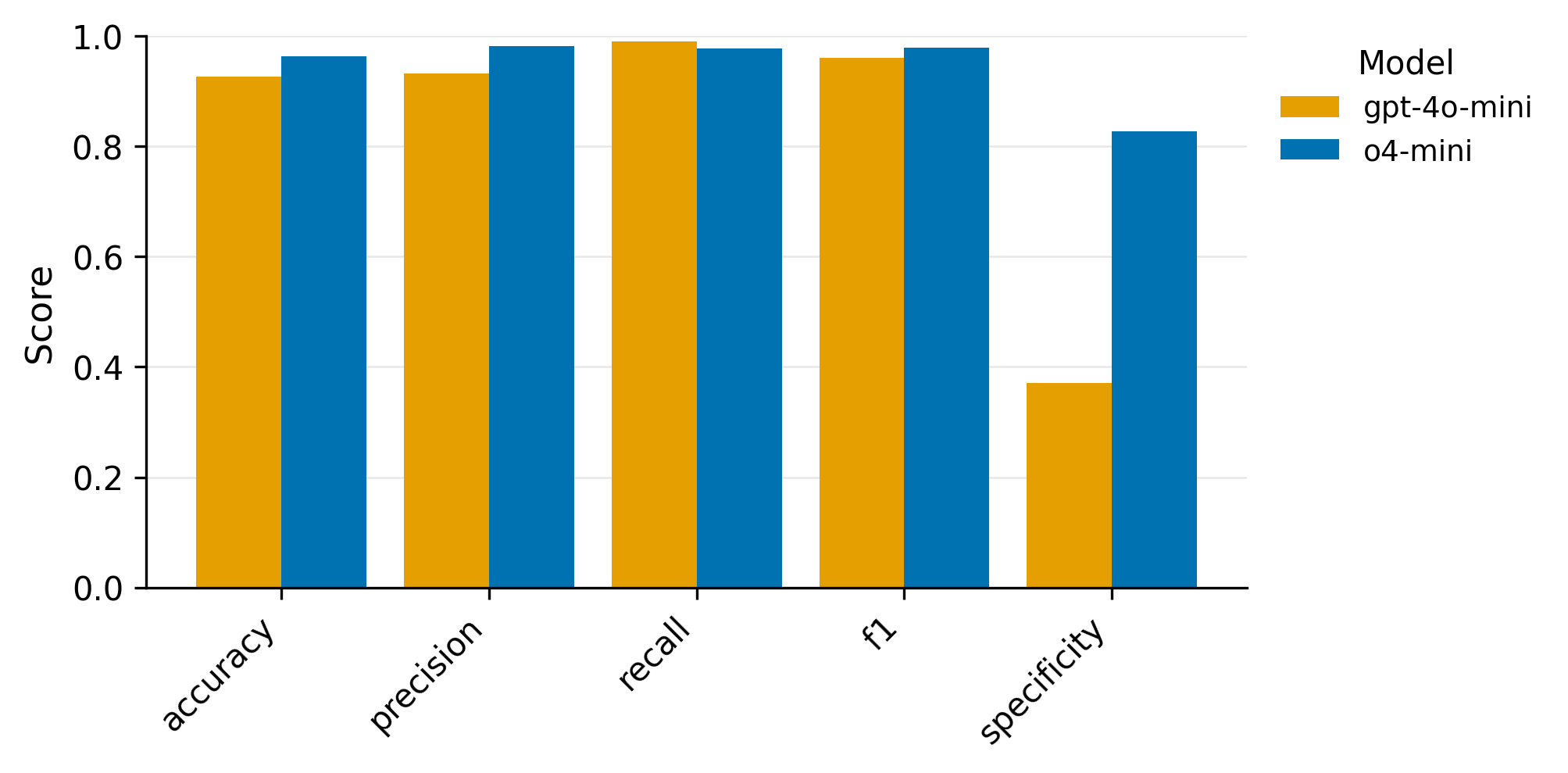}
\caption{Full-text evidence direction performance (\psthree\ vs.\ \bsthree) on successfully matched examples. The reasoning model (\texttt{o4-mini}) substantially improves specificity while maintaining high recall.}
\label{fig:ps3-bs3-metrics}
\end{figure}

Figure~\ref{fig:ps3-bs3-metrics} (Appendix Table~\ref{tab:fulltext-direction}) reports direction classification (\psthree\ vs.\ \bsthree) on successfully matched examples, with \texttt{not\_clear} summarized through coverage. Both models performed strongly on decided cases. Notably, \texttt{o4-mini} achieved much higher specificity (0.828 vs.\ 0.371), indicating fewer false \psthree\ calls when \bsthree\ was the ground truth. This improvement coincided with lower coverage (0.916 vs.\ 0.994), consistent with more conservative no-decision behavior under ambiguity.

\subsection{Evidence strength and joint direction+strength remain challenging}
\begin{figure}[t]
\centering
\includegraphics[width=0.92\linewidth]{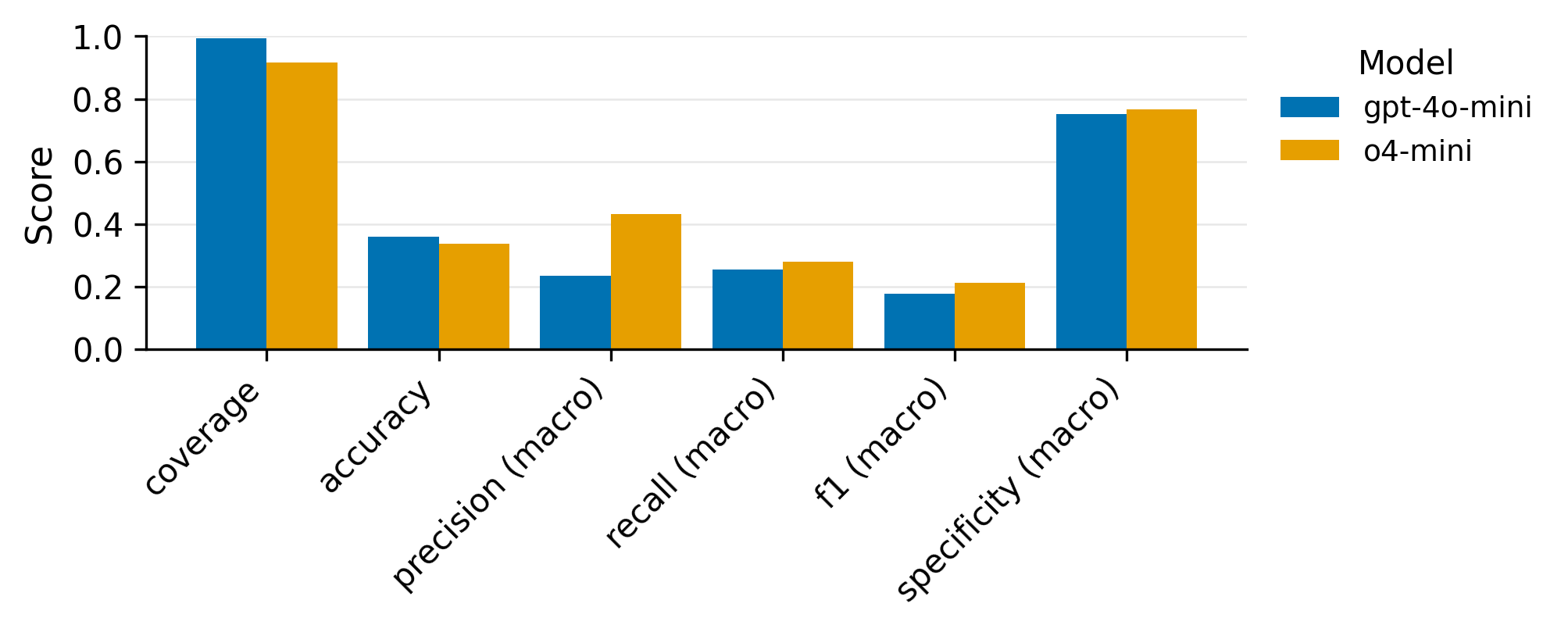}
\caption{Evidence strength classification performance (4-way) on successfully matched examples. Strength grading remains challenging for both models.}
\label{fig:strength-metrics}
\end{figure}

\begin{figure}[t]
\centering
\includegraphics[width=0.92\linewidth]{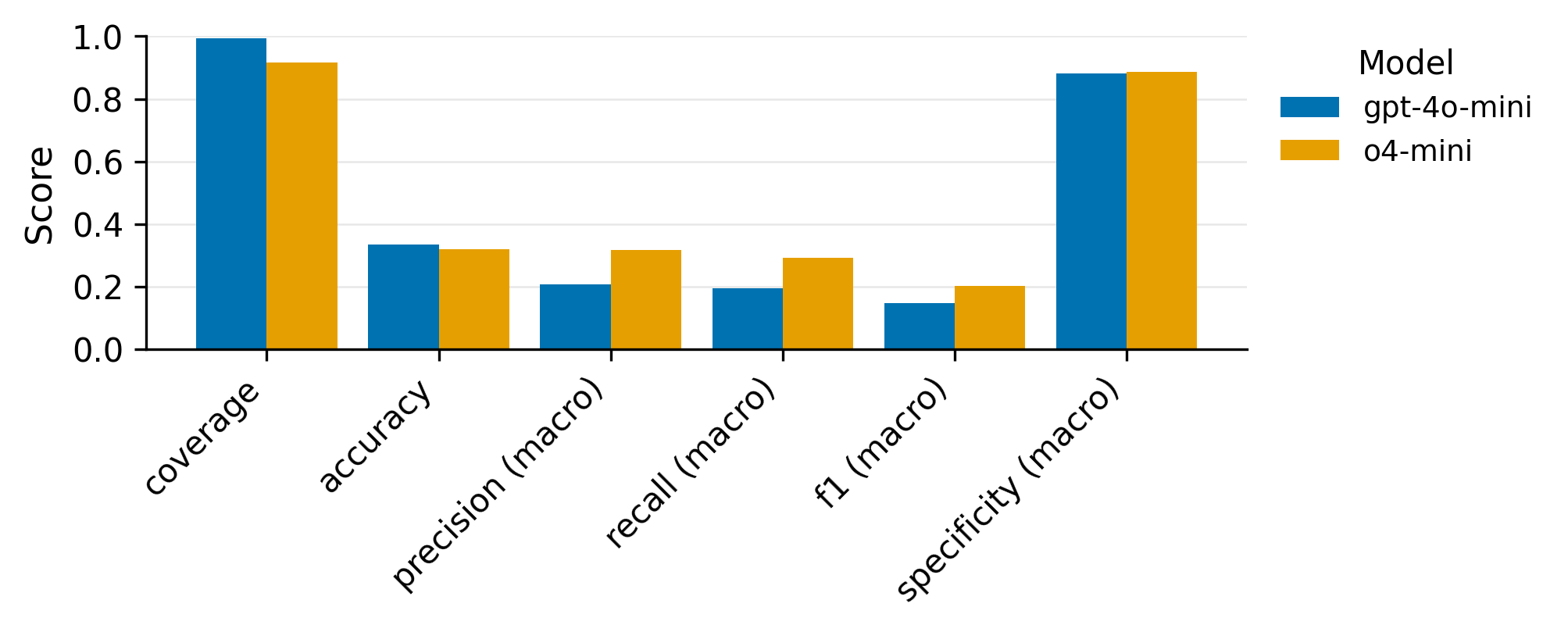}
\caption{Joint direction+strength classification performance (8-way) on successfully matched examples.}
\label{fig:joint-metrics}
\end{figure}

Strength grading remained difficult. As shown in Figure~\ref{fig:strength-metrics} (Appendix Table~\ref{tab:fulltext-strength}), both models achieved low accuracy ($\approx 0.34$--$0.36$) and low macro-F1 ($\approx 0.18$--$0.21$) for 4-way strength prediction (supporting, moderate, strong, very strong). The joint direction+strength task (Figure~\ref{fig:joint-metrics}, Appendix Table~\ref{tab:fulltext-joint}) was similarly challenging. These results suggest that while overall functional direction is often recoverable, strength depends on assay validation and calibration signals that may be implicit, fragmented, or relegated to figures, tables, or supplements.

\subsection{LLM-as-judge evaluation of evidence summary correspondence}
\begin{figure}[t]
\centering
\includegraphics[width=0.92\linewidth]{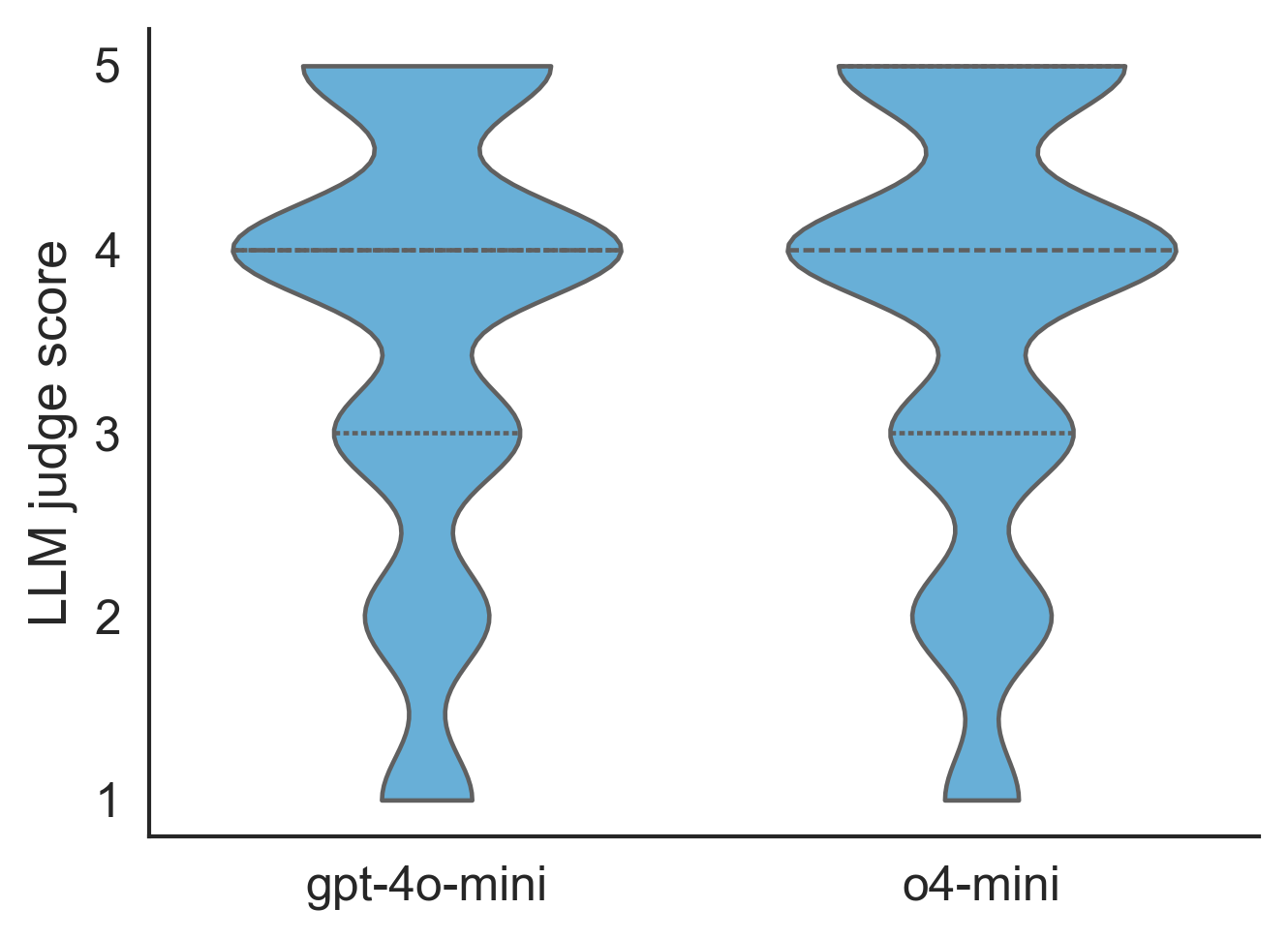}
\caption{LLM-as-judge correspondence scores (1--5) comparing LLM-generated evidence summaries to ClinGen \psthree/\bsthree rationales. Violin plots show the distribution over evaluated examples; higher scores indicate stronger factual alignment and better coverage of the key experimental conclusions.}
\label{fig:llm-judge-scores}
\end{figure}

\begin{figure}[t]
\centering
\includegraphics[width=0.92\linewidth]{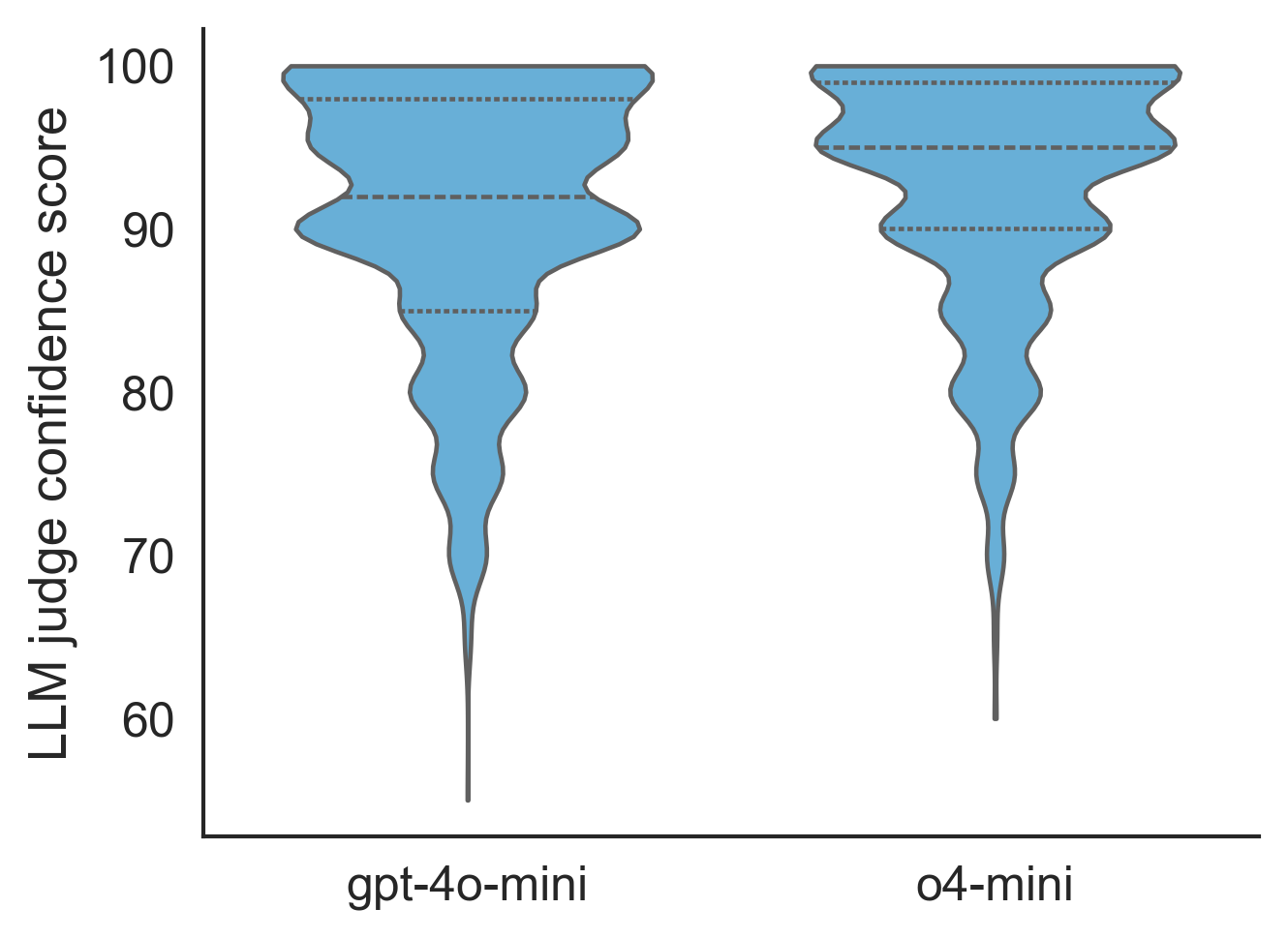}
\caption{LLM judge confidence scores (0--100) associated with the correspondence assessments in Figure~\ref{fig:llm-judge-scores}. Higher values indicate greater judge certainty in the assigned correspondence score.}
\label{fig:llm-judge-confidences}
\end{figure}

Beyond label accuracy, we assessed whether each model’s extracted evidence summary matched the expert-written ClinGen \psthree/\bsthree rationale using an independent LLM rater. The judge assigned a 1--5 correspondence score (higher = better alignment), reflecting factual consistency and coverage of the key experimental outcomes, and provided a calibrated confidence score (0--100) for each assessment. Figure~\ref{fig:llm-judge-scores} shows the distribution of correspondence scores, and Figure~\ref{fig:llm-judge-confidences} summarizes the score confidence distribution. We observed comparable score distributions for both models; however, the confidence distribution was higher for \texttt{o4-mini} than for \texttt{gpt-4o-mini}. Together, these distributions characterize summary faithfulness and perceived reliability, complementing direction/strength classification metrics.

\subsection{End-to-end pipeline output and report visualization}

While the benchmark isolates individual sub-tasks (abstract screening, variant matching, and full-text evidence interpretation), real curation workflows require stitching these steps into a traceable end-to-end process. We therefore implemented \toolname, a variant-centric system that begins with a genomic variant and produces a curator-oriented evidence package aligned to \acmg functional criteria.

Given a variant specified by coordinates (chromosome, position, reference, alternate), \toolname first normalizes the variant into a consolidated identifier set (e.g., rsID when available; HGVSg/HGVSc/HGVSp; gene symbol; transcript context) and expands synonyms to reflect how variants are typically referenced across papers. It then performs literature retrieval (via LitVar2), screens candidate abstracts for likely variant-linked functional assays, attempts PDF acquisition, and applies multimodal extraction on the retained full papers. For each matched paper, the system extracts structured experiment records (assay, system, readout, comparator, direction, controls, caveats, and supporting locations) and then integrates the evidence across papers into an \acmg-aligned assessment. Importantly, \toolname is designed to be conservative under uncertainty: when variant identity cannot be confidently aligned or functional direction/strength cannot be supported from the text and figures, the pipeline returns a no-decision outcome rather than forcing a \psthree/\bsthree call.

The primary output is a human-reviewable HTML or PDF report intended to support rapid curator verification. The report consolidates: (i) normalized variant metadata and transcript context; (ii) retrieved literature with abstract-screening outcomes and download status; (iii) per-paper variant matching notes and extracted experiments with citations to the originating paper sections/figures when available; and (iv) a final integrated \acmg-aligned summary that highlights direction, proposed strength, confidence, and key considerations that may affect \psthree/\bsthree use (e.g., assay validation signals, calibration controls, and concordance with disease mechanism). In addition to the report, \toolname can return a structured JSON record to support downstream programmatic analyses.

\begin{figure}[t]
\centering

\includegraphics[width=\linewidth]{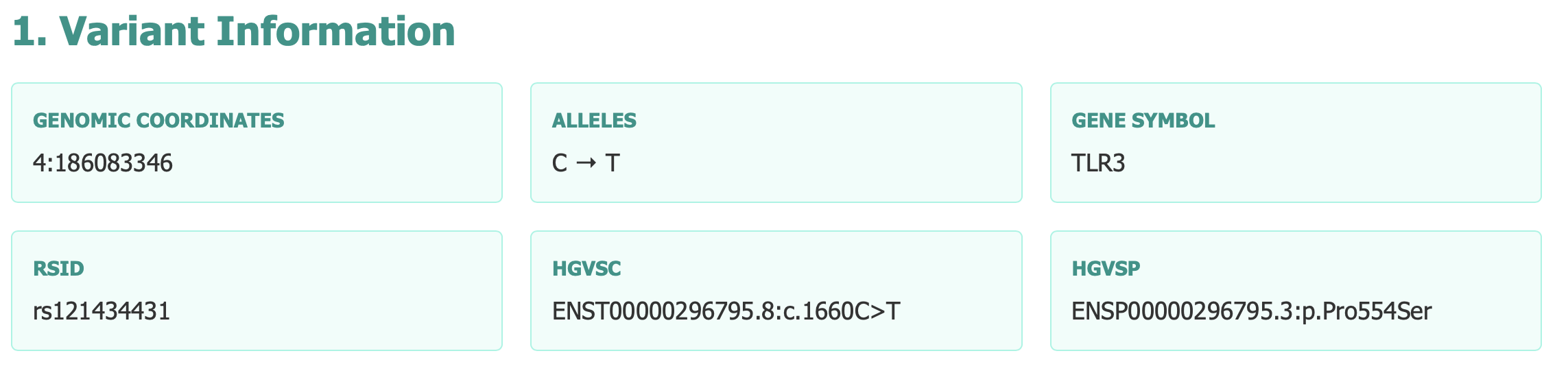}

\vspace{1em}

\includegraphics[width=\linewidth]{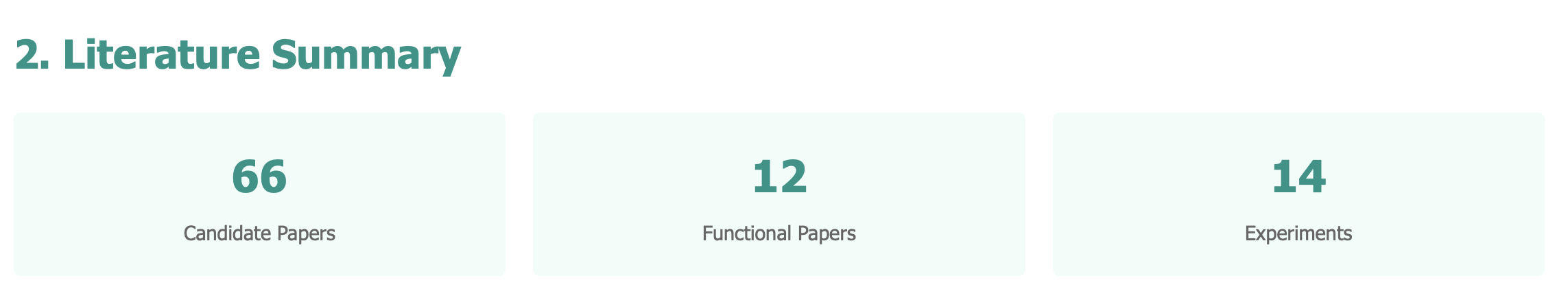}

\vspace{1em}

\includegraphics[width=\linewidth]{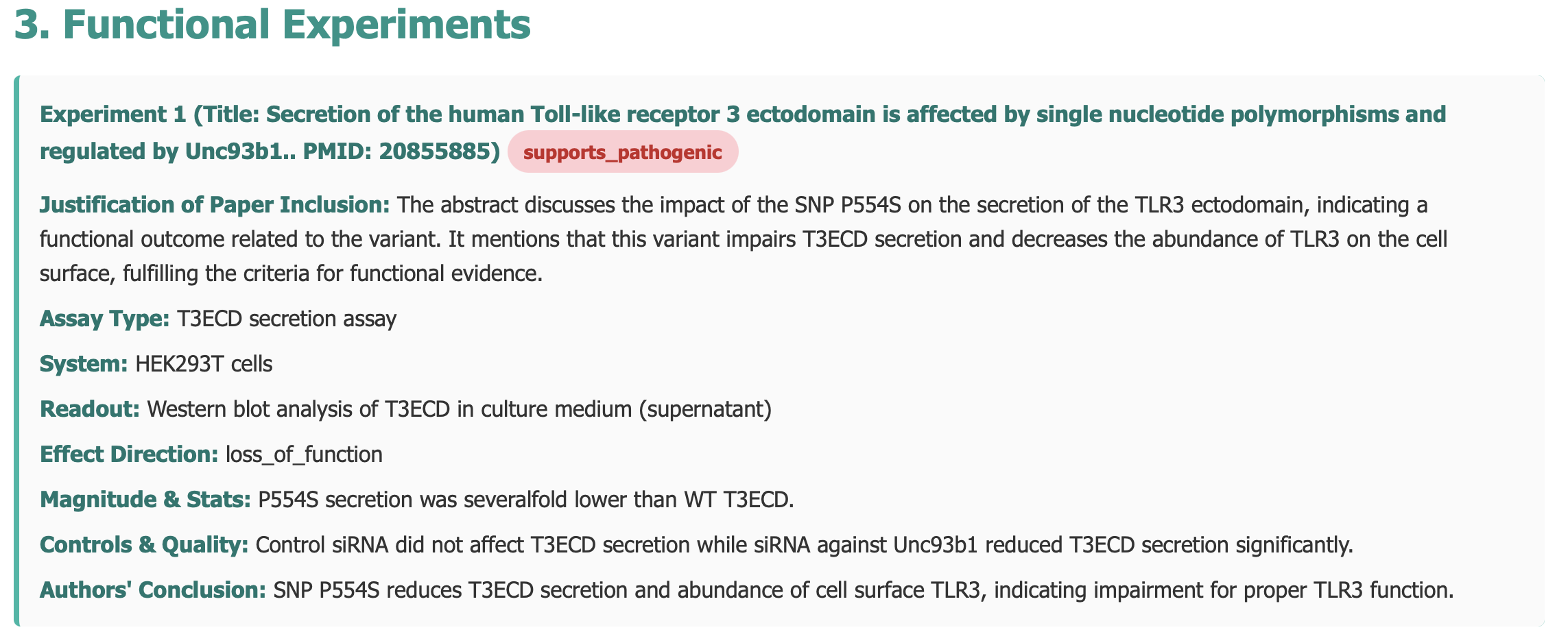}

\vspace{1em}

\includegraphics[width=\linewidth]{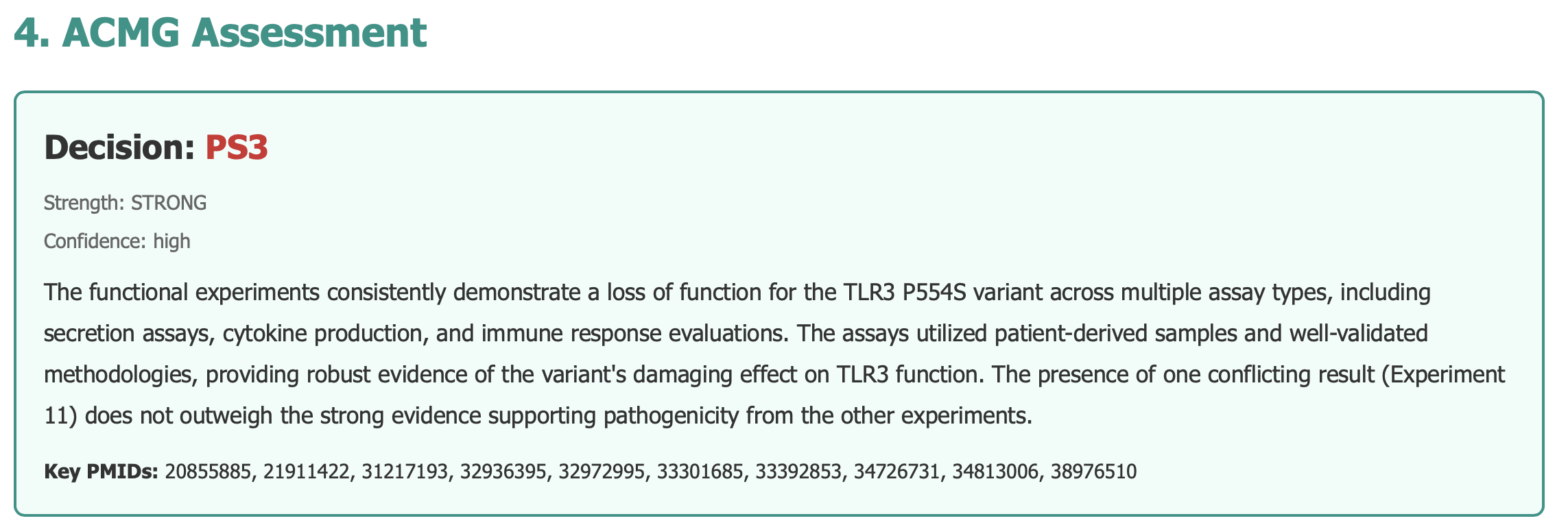}

\caption{Representative screenshots from the \toolname evidence report. The report is designed for curator review and compiles variant normalization and transcript context, retrieved literature with screening decisions, per-paper variant matching and extracted functional experiments, and an integrated \acmg-aligned assessment summarizing \psthree/\bsthree direction, proposed strength, confidence, and key considerations.}
\label{fig:report-screens}
\end{figure}

Figure~\ref{fig:report-screens} shows representative sections of a generated report, including (top to bottom) variant normalization, candidate-paper screening and selection, extracted functional experiments, and the integrated curator-facing assessment.

\section{Discussion}

We evaluated how far modern multimodal LLMs can support functional-evidence curation in practice, spanning study identification, variant-linked extraction from full paper, and \psthree/\bsthree-aligned interpretation. Across tasks, a consistent picture emerged: abstract screening is already useful as a sensitive filter, full-paper \psthree\ versus \bsthree\ direction is largely tractable \emph{once variant identity is confidently established}, and evidence \emph{strength} is the most fragile component because it relies on assay validation and calibration signals that are frequently implicit, distributed across figures and methods, or omitted from the main PDF. This motivates richer document-understanding systems. To this end, \toolname includes an optional tool-augmented, agentic extraction mode designed to better surface evidence embedded in visual elements.

At the abstract stage, both reasoning and non-reasoning models achieved high recall (0.885--0.904), supporting the use of LLMs as a first-pass filter to prioritize likely functional studies for curator review (Figure~\ref{fig:abstract}; Table~\ref{tab:abstract-highrecall}). In this workflow, recall is the dominant objective: false positives primarily increase reviewer workload, whereas false negatives risk hiding clinically relevant experiments. The remaining errors are consistent with the information bottleneck in abstracts, which often omit whether variants were directly tested or report broad biological conclusions without sufficient experimental specificity.

For full-paper processing, our results emphasize that variant identity alignment is a prerequisite rather than a sub-detail. A substantial fraction of paper--variant pairs failed matching (26--30\%; Figure~\ref{fig:variant-matching}; Table~\ref{tab:variant-matching}), reflecting heterogeneous nomenclature and incomplete identifiers in manuscripts. Because misattribution of results to the wrong variant is a high-risk failure mode, we treated matching as an explicit gate and preferred conservative abstention under uncertainty. Conditional on successful matching, both models performed strongly on direction classification, and the reasoning model (\texttt{o4-mini}) achieved markedly higher specificity (0.828 vs.\ 0.371) while maintaining high recall (Figure~\ref{fig:ps3-bs3-metrics}; Table~\ref{tab:fulltext-direction}). Practically, this corresponds to fewer erroneous \psthree\ calls on \bsthree-ground-truth cases, a desirable behavior when the downstream implication is that damaging functional impact is being asserted. The gain in specificity came with reduced coverage (0.916 vs.\ 0.994), consistent with a more conservative tendency to return \texttt{not\_clear} when evidence is mixed, weakly supported, or poorly anchored to the target variant.

In contrast, strength grading and the joint direction+strength task remained challenging (Figures~\ref{fig:strength-metrics} and~\ref{fig:joint-metrics}; Tables~\ref{tab:fulltext-strength} and~\ref{tab:fulltext-joint}). This gap is expected given how ClinGen recommendations \citep{Brnich2019-zm} operationalize strength: strength is not simply the magnitude of an observed effect, but depends on assay validity (controls, calibration, replication, dynamic range), concordance with disease mechanism, and the extent to which results generalize across variants and systems. These properties are often conveyed through figures, supplementary materials, and methodological nuance rather than a single extractable statement in the main text. Together, the direction and strength findings support an assistive deployment model: LLMs can extract structured experimental evidence and propose direction decisions with traceable excerpts, while strength should be treated as a curator-auditable hypothesis grounded in explicitly surfaced validation signals.

The LLM-as-judge analysis complements label-based metrics by asking whether extracted summaries preserve the substance of expert rationales. We observed comparable correspondence score distributions across models, with higher judge confidence for \texttt{o4-mini} (Figures~\ref{fig:llm-judge-scores} and~\ref{fig:llm-judge-confidences}). Interpreted cautiously, this suggests that both models can often produce summaries aligned with curator narratives, while the reasoning model’s outputs may be perceived as more consistently evaluable by an independent rater. This further supports using LLM outputs as curated drafts that accelerate review, while still requiring verification and explicit provenance.

We operationalize these insights in \toolname, an end-to-end open-source workflow designed around curator needs: aggressive synonym expansion and retrieval, high-recall abstract filtering, conservative variant matching, structured experiment extraction, and report generation that emphasizes auditable evidence and decision rationale (Figure~\ref{fig:report-screens}). In this framework, LLMs provide the most value when they compress reading and organization time, while deferring high-stakes adjudication to human review when uncertainty is high.

Several limitations motivate concrete next steps. First, we evaluated only two multimodal LLMs; broader benchmarking across additional model families and sizes is needed to assess generality and to characterize whether observed trade-offs (e.g., specificity versus coverage) hold across architectures and training regimes. Second, while \toolname includes an optional agentic extraction mode intended to better capture information in figures and tables, we did not quantitatively evaluate its accuracy benefit due to computational and time constraints. We therefore frame it as an exploratory component and a clear direction for future work. Third, the main PDF is often insufficient for strength grading and sometimes even for direction, because validation details and calibration controls are frequently reported in supplementary files, extended methods, or external repositories. Treating supplements as first-class inputs and explicitly tracking whether claims are supported by main-text versus supplementary evidence are important pipeline extensions. Fourth, the benchmark inherits uncertainty from underlying curated rationales: ClinGen summaries reflect expert interpretation, but judgments can vary across curators, panels, and time as standards evolve. Future work could quantify inter-curator variability where multiple rationales exist and evaluate against multi-annotator references or adjudicated consensus to better separate model error from label ambiguity. Finally, although \toolname emphasizes auditable extraction, LLMs can still generate unsupported inferences. Strengthening provenance guarantees is therefore a priority, including tighter evidence-first constraints, automated consistency checks that flag claims lacking textual or figure support, and robustness testing via systematic perturbations (prompt variants, formatting changes, alternative schemas) within a versioned evaluation harness.

In summary, multimodal LLMs offer a compelling opportunity to make functional-evidence curation both faster and more consistent. They can absorb much of the upfront effort by organizing relevant information into a structured, curator-friendly form and producing draft rationales that are straightforward to review. By shifting effort from information gathering to focused verification, this approach can help standardize how evidence is organized and compared across studies and variants, enabling higher-throughput activities such as panel updates, large-scale re-interpretation, and ongoing literature surveillance while keeping final adjudication with curators. In this capacity, \toolname serves as an extensible foundation for incorporating additional evidence sources and reliability checks, with the goal of reducing time-to-curation without compromising transparency.

\section{Data and Code Availability}
The benchmark data and \toolname\ pipeline code are available respectively at \url{https://github.com/AliSaadatV/LLM_func_exp_bench} and \url{https://github.com/AliSaadatV/AcmGENTIC}.

\bibliographystyle{oup-abbrvnat}
\bibliography{reference}

\begin{appendices}

\FloatBarrier
\section{Tables}\label{tables}

\begin{table}[t]
\caption{Abstract-level functional experiment detection.}
\label{tab:abstract-highrecall}
\begin{tabular}{l S[table-format=1.4] S[table-format=1.4] S[table-format=1.4] S[table-format=1.4] S[table-format=1.4] S[table-format=1.4] S[table-format=1.4] S[table-format=1.4]}
\toprule
Model & {Accuracy} & {Precision} & {Recall} & {F1} & {Specificity}  \\
\midrule
\texttt{gpt-4o-mini} & 0.7467 & 0.6878 & 0.9036 & 0.7810 & 0.5898  \\
\texttt{o4-mini}     & 0.7665 & 0.7156 & 0.8847 & 0.7912 & 0.6484  \\
\bottomrule
\end{tabular}
\end{table}

\begin{table}[t]
\caption{Variant matching status counts for full-paper processing.}
\label{tab:variant-matching}
\begin{tabular}{l rr}
\toprule
Variant Match Status & \texttt{gpt-4o-mini} & \texttt{o4-mini} \\
\midrule
\texttt{matched} & 283 & 278 \\
\texttt{heuristic\_matching} & 56 & 42 \\
\texttt{single\_variant\_study\_matching} & 0 & 3 \\
\texttt{variant\_matching\_unsuccessful} & 121 & 137 \\
\texttt{NA} & 3 & 3 \\
\midrule
\textbf{Total successfully matched} & 339 & 323 \\
\bottomrule
\end{tabular}
\end{table}

\begin{table}[t]
\caption{Full-text evidence direction classification (\psthree\ vs.\ \bsthree) conditional on successful variant matching. \texttt{not\_clear} is treated as a no-decision outcome and summarized by coverage.}
\label{tab:fulltext-direction}
\begin{tabular}{l r r S[table-format=1.4] S[table-format=1.4] S[table-format=1.4] S[table-format=1.4] S[table-format=1.4] S[table-format=1.4] S[table-format=1.4] S[table-format=1.4]}
\toprule
Model &  {Coverage} & {Accuracy} & {Precision} & {Recall} & {F1} & {Specificity}  \\
\midrule
\texttt{gpt-4o-mini} & 0.9941 & 0.9258 & 0.9315 & 0.9901 & 0.9599 & 0.3714  \\
\texttt{o4-mini}    & 0.9164 & 0.9628 & 0.9812 & 0.9775 & 0.9794 & 0.8276  \\
\bottomrule
\end{tabular}
\end{table}

\begin{table}[t]
\caption{Full-text evidence strength classification (4-way), conditional on successful variant matching; \texttt{not\_clear} treated as a no-decision outcome (coverage reported).}
\label{tab:fulltext-strength}
\begin{tabular}{l r r S[table-format=1.4] S[table-format=1.4] S[table-format=1.4] S[table-format=1.4] S[table-format=1.4] S[table-format=1.4]}
\toprule
Model  & {Coverage} & {Accuracy} & {Prec$_{\text{macro}}$} & {Rec$_{\text{macro}}$} & {F1$_{\text{macro}}$} & {Spec$_{\text{macro}}$} \\
\midrule
\texttt{gpt-4o-mini} &  0.9941 & 0.3591 & 0.2363 & 0.2555 & 0.1766 & 0.7520 \\
\texttt{o4-mini}     &  0.9164 & 0.3378 & 0.4318 & 0.2802 & 0.2115 & 0.7654 \\
\bottomrule
\end{tabular}
\end{table}

\begin{table}[t]
\caption{Joint evidence direction+strength classification (8-way), conditional on successful matching; \texttt{not\_clear} treated as a no-decision outcome.}
\label{tab:fulltext-joint}
\begin{tabular}{l r r S[table-format=1.4] S[table-format=1.4] S[table-format=1.4] S[table-format=1.4] S[table-format=1.4]}
\toprule
Model & {Coverage} & {Accuracy} & {Prec$_{\text{macro}}$} & {Rec$_{\text{macro}}$} & {F1$_{\text{macro}}$} \\
\midrule
\texttt{gpt-4o-mini} & 0.9941 & 0.3353 & 0.2068 & 0.1947 & 0.1473 \\
\texttt{o4-mini}     & 0.9164 & 0.3209 & 0.3181 & 0.2915 & 0.2022 \\
\bottomrule
\end{tabular}
\end{table}

\FloatBarrier
\section{Prompt Templates}\label{app:prompts}

\subsection{ClinGen curator summary parsing prompt}
\begin{lstlisting}[caption=]
You are assisting with ACMG/AMP functional evidence extraction.

Input text (ClinGen curation summary):
---
{summary_text}
---

From the above text, extract the following strictly structured JSON:

{
  "PS3_pmids": [... PMIDs as integers ...],
  "BS3_pmids": [... PMIDs as integers ...],
  "PS3_level": "PS3" | "PS3_moderate" | "PS3_supporting" | "none",
  "BS3_level": "BS3" | "BS3_moderate" | "BS3_supporting" | "none",
  "PS3_comments": "comments" | "none",
  "BS3_comments": "comments" | "none"
}

Rules:
- A PMID must be a real integer.
- If no papers exist, output an empty list.
- If levels or comments are not present, output null or omit the key.
\end{lstlisting}

\subsection{Abstract screening prompt}
\begin{lstlisting}[caption=]
You are a clinical variant interpretation curator performing an 
abstract-level screen for ACMG/AMP PS3/BS3 relevance.

Goal: Decide ONLY whether the abstract contains ANY experimental 
(wet-lab) functional evidence about the effect of one or more 
genetic variants/mutations/alleles/mutants on gene product function.

Bias / sensitivity requirement (important):
This screen is intentionally high-sensitivity. If there is 
reasonable doubt, classify as 1 so the paper can be reviewed 
downstream. Default to 1 whenever BOTH (i) variant/mutant language 
and (ii) any wet-lab functional assay signal are present.

Classify functional_experiment = 1 if the abstract shows BOTH:
A) Variant-or-mutant subject (broad; exact IDs NOT required)
B) Wet-lab functional assay + outcome statement

Count as functional evidence if the abstract includes:
1) Protein/biochemical function (activity, binding, stability, etc.)
2) Cell-based functional consequences (reporter, rescue, etc.)
3) RNA-level functional assays (splicing, NMD, etc.)
4) Model systems with variant-level manipulation
5) Patient-derived functional assays

Return functional_experiment = 0 ONLY when it is clearly NOT 
functional variant testing (pure in silico, genetic association, 
or gene biology without variant testing).
\end{lstlisting}

\subsection{Full-paper experiment extraction and PS3/BS3 classification prompt}
Due to length, the full prompt is provided in the code repository (\url{https://github.com/AliSaadatV/AcmGENTIC}). Key sections include:
\begin{itemize}
\item \textbf{Variant Matching (Soft Gate)}: Instructions for building an equivalents set and matching via strict, heuristic, or single-variant-study criteria.
\item \textbf{Experiment Extraction}: Structured schema for assay, system, readout, comparator, result, controls, and caveats.
\item \textbf{PS3/BS3/not\_clear Assignment}: Definitions and strength criteria aligned with ClinGen SVI recommendations.
\end{itemize}

\subsection{Evidence integration prompt}
\begin{lstlisting}[caption=]
You are a clinical variant interpretation curator specializing in 
ACMG/AMP PS3/BS3 functional evidence criteria.

ACMG/AMP DEFINITIONS:
- PS3: Well-established functional studies supportive of DAMAGING effect
- BS3: Well-established functional studies show NO DAMAGING effect

ClinGen SVI STRENGTH FRAMEWORK:
- very_strong: Rigorous statistical validation, OddsPath calculation
- strong: Multiple independent, well-validated assays, 2+ concordant papers
- moderate: Validated assay with 11+ variant controls
- supporting: Basic assay with limited validation

DECISION RULES:
1. PS3: Consistently abnormal function, consistent with disease mechanism
2. BS3: Consistently normal function, adequate assay coverage
3. not_clear: Conflicting, insufficient, or ambiguous evidence

Return JSON: {decision, strength, confidence, narrative, 
experiment_evaluations, key_considerations}
\end{lstlisting}

\section{Pipeline Usage Examples}\label{app:usage}

\subsection{Command-line interface usage}
\begin{lstlisting}[language=bash,caption=]
# Basic analysis with HTML+PDF output
python main.py --chrom 2 --pos 162279995 --ref C --alt G

# With custom model and agentic extraction
python main.py --chrom 2 --pos 162279995 --ref C --alt G \
    --model gpt-4o --agentic

# Non-interactive mode
python main.py --chrom 17 --pos 41244694 --ref T --alt A \
    --no-interactive --output-format dict > results.json
\end{lstlisting}

\subsection{Python API usage}
\begin{lstlisting}[language=Python,caption=]
from main import analyze_variant

result = analyze_variant(
    chrom="2",
    pos=162279995,
    ref="C",
    alt="G",
    assembly="GRCh38",
    pdf_path="papers",
    download_pdfs=True,
    interactive=False,
)

# Access results
decision = result["assessment"]["decision"]  # "PS3", "BS3", or "none"
strength = result["assessment"]["strength"]
narrative = result["assessment"]["narrative"]
\end{lstlisting}

\end{appendices}

\end{document}